\documentclass[a4paper,12pt]{article}

\usepackage[margin=1in]{geometry}

\usepackage{hyperref}

\hypersetup{
	colorlinks=true,
	linkcolor=blue,
	filecolor=magenta,
	urlcolor=cyan,
}

\usepackage{amsmath} %for align
\usepackage{graphicx}
\usepackage{subcaption}
\usepackage{pxfonts}
\usepackage{lineno}% if you use PostScript figures in your article
\usepackage{amssymb}
\usepackage{footnote}%!!!
\usepackage{multirow}
\usepackage{varwidth}
\usepackage{array}
\usepackage{epstopdf}
\usepackage{color} %for \textcolor and \rr to work

\usepackage{grffile} %support of dots in image file names
\usepackage{makecell} % allows to split the table cell into two lines
\usepackage[backend=bibtex,url=true,sorting=none,doi=true,maxnames=3,firstinits=true,citestyle=numeric-comp]{biblatex}

\bibliography{library}

\usepackage{lscape}
\usepackage{authblk}

\usepackage{caption}
\DeclareCaptionFormat{myformat}{#1 (color online)#2#3}
{\captionsetup{format=myformat,labelsep=colon}
	\figure
}%
{\endfigure}%

\renewbibmacro*{doi+eprint+url}{%
	\iftoggle{bbx:doi}
	{\printfield{doi}}
	{}
	\newunit\newblock
	\iftoggle{bbx:url}
	{\iffieldundef{doi}{
			\iffieldundef{arxivId}{	\printfield{url}}{}}{}
	}
	{}%
	\newunit\newblock
	\iftoggle{bbx:eprint}
	{\usebibmacro{eprint}}
	{}%
}

\title{ROSA: Reconstruction of Spectrogram Autocorrelation for  Self-Amplified Spontaneous Emission Free-Electron Lasers}

\author[1]{Svitozar Serkez}
\author[2,3]{Oleg Gorobtsov}
\author[4]{Bohdana Sobko}
\author[1]{Natalia Gerasimova}
\author[1]{Gianluca Geloni}

\affil[1]{European XFEL GmbH, Hamburg, Germany}
\affil[2]{Deutsches Elektronen-Synchrotron (DESY), Hamburg, Germany}
\affil[3]{Present address: Department of Materials Science and Engineering, Cornell University, Ithaca, NY, USA}%14850
\affil[4]{Lviv National University, Lviv, Ukraine}

\begin{document}

\maketitle

%\begin{center}
%    \Large{\rr{DRAFT}}
%\end{center}

\begin{abstract}
In recent years X-ray Free Electron Lasers (XFELs) proved to be unmatched sources of ultrashort pulses of spatially coherent quasimonochromatic X-ray radiation.
Diagnostics of XFEL emission properties, in particular pulse duration, spectrum and temporal profile is extremely important in order to analyze the experimental results.
In this paper we propose a cost-effective method to examine these properties of Self-Amplified Spontaneous Emission (SASE) pulses.
It only requires an ensemble of measured SASE spectra and provides the temporal autocorrelation of the ensemble-averaged Wigner distribution of SASE FEL pulses.
\end{abstract}

% \end{frontmatter}

% main text

\section{Introduction}

Free-electron laser sources provide high power, ultrashort  pulses of narrow-bandwidth radiation at tunable wavelength~\cite{Saldin2010book, McNeil2010, Pellegrini2016}. 

In essence, a high-gain free electron laser is a combination of accelerator that produces a high-quality ultrarelativistic electron beam and undulator - a periodic magnetic structure where that beam is forced to oscillate transversely with respect to the main direction of motion.
A positive feedback loop is formed in the undulator where electromagnetic field is amplified in an exponential fashion. A linear combination of electron density modulation, energy modulation and electromagnetic field present at the target wavelength provide initial conditions for the FEL process.
Energy transfer from the electron beam to the electromagnetic radiation pulse is sustained when the resonance condition

\begin{equation}
\lambda = \lambda_u \frac{ 1 + K^2/2}{2 \gamma^2}
\label{eq:fel}
\end{equation}

is satisfied, where  $ \lambda_u $ is the undulator period, $ \gamma $ is the Lorenz factor of the reference electron and $ K=eH\lambda_u / 2\pi m_ec^2 $ is the undulator parameter, $e$ is an electron charge, $m_e$ - the electron rest mass, $H$ - the maximum on-axis magnetic field  and $c$ - a speed of light in vacuum. 

Note that if the average electron energy at the head of the beam differs significantly from that of the tail, the undulator resonance condition varies from head to tail, and the resulting radiation pulse is frequency chirped. We also remark that ideally, an electron beam should have a minimum energy spread and energy chirp to maximize the FEL gain and minimize the FEL bandwidth. Therefore, a monoenergetic electron beam is preferable to generate FEL radiation.

The electron beam phase space and the radiation output characteristics are closely related and difficult to diagnose. 

X-band transverse deflecting radio-frequency cavities (XTCAV)~\cite{Behrens2014} are powerful tools to directly measure the longitudinal phase space of FEL-class electron beams, allowing for the determination of the overall energy chirp, i.e. the correlated energy spread in the electron beam in order to minimize it. 
When installed downstream the FEL undulator system, they can also be used as diagnostics tools for the FEL pulse, allowing for calculating the pulse duration and even its spectrogram.

When the longitudinal phase space of an electron beam cannot be measured, it may be hard to optimize the accelerator parameters for lasing, e.g. for minimizing the radiation pulse duration. In this case one may resort to an analysis of the SASE FEL spectra. 
For example, based on the statistical properties of the radiation~\cite{Saldin1998} one may apply spectral correlation analysis and estimate the average duration of the SASE FEL pulse assuming a certain temporal power profile~\cite{Lutman2012, Inubushi2012}, further tuning the accelerator to decrease such duration by effect. 
However, the close relation between electron phase space and radiation characteristics must always be taken into account in this process. 
For example, it was shown in~\cite{Krinsky2003} that the chirp in an electron beam affects the range of spectral coherence, and hence the spectrum-based estimation of the SASE pulse duration. 

Here we present a method to estimate duration and shape of SASE radiation pulses based on the reconstruction of autocorrelation of the ensemble-averaged pulse Wigner distribution.
In Section~\ref{sec:theory} we discuss statistical properties of the SASE radiation, and show how the Wigner distribution autocorrelation can be calculated from measured SASE FEL spectra.
In Section~\ref{sec:simulations} we present SASE FEL spectra simulated with the FEL code GENESIS~\cite{GENE} assuming different electron beam parameters and we compare results of the reconstruction with known pulse shapes.
Finally, in Section~\ref{sec:conclusions} we come to conclusions.

\section{Theory}\label{sec:theory}

In this section we describe the theoretical background at the basis of our retrieval algorithm. We fix our notations and give several definitions. We then proceed to the formulation of the Wigner distribution autocorrelation reconstruction algorithm. 
%Subsequently, we demonstrate the relation between the retrieved function and the average duration of the FEL photon pulse. We perform this last step based on the assumption of Gaussian statistics for the FEL pulses, meaning that the theory is valid, strictly speaking, only in the linear regime. However, as we will see from numerical simulations in the following sections, even at saturation, where Gaussian statistics does not strictly apply, the application of the Sievert relation does not yield large errors in the reconstruction of the radiation pulse length. In fact, such error will be found to be practically negligible compared to uncertainties due to machine fluctuations, which are not related with SASE stochastic process. We will also see, however, that these last uncertainties can be kept under control too, by properly selecting the data to analyze.

%One does not simply walk into Mordor Why not?

\subsection{Definitions and conventions}\label{sec:definitions}

%Correlation analysis of the ...
Consider a scalar field $E(t)$ in the time domain and its Fourier transform $\bar{E}(\omega)$
\begin{align}
\bar{E}(\omega) &= \frac{1}{2\pi} \int_{-\infty}^{\infty} dt~ E(t) \exp(i \omega t) \ ,\cr
E(t) &= \int_{-\infty}^{\infty} d\omega~ \bar{E}(\omega) \exp(-i \omega t) \ .
\label{FT1}
\end{align}
% OG Th p29, MW p56

We introduce the slowly varying field amplitude $\widetilde{E}(\omega)$ through the relation

\begin{equation}
\bar{E}(\omega) = \widetilde{E}(z, \omega) \exp(i \omega_c z/c) \ , %\mathrm{\rr{check minus in the exponent}}
\label{slow}
\end{equation}
where $ \omega_c $ is a central, carrier wavelength and $z$ is the direction of propagation.
Similarly to radiation power, measured single-shot spectra are proportional to the square-modulus of the single-shot slowly varying envelope 
\begin{align}
I(t) &\equiv  E(t)E^*(t),\cr
\widetilde{I}(\omega) &\equiv  \widetilde{E}(\omega)\widetilde{E}^*(\omega) \ .
\label{singlespec}
\end{align}

The statistical  autocorrelation function of the field $ E(t) $ in the time domain can then be defined as\footnote{Note that the autocorrelation function depends on both time $ t $ and time separation $ \Delta t $, allowing to describe non-stationary radiation fields.}
% Saldin physics p148, Goodman p73
\begin{equation}
\Gamma(t, \Delta t) = \left\langle {E}\left(t-\frac{\Delta t}{2}\right) {E}^*\left(t+\frac{\Delta t}{2}\right)\right\rangle,
\label{G1deft}
\end{equation}
where angle brackets $\langle\rangle$ denote ensemble average.
The intensity autocorrelation function is, instead
\begin{equation}
\Gamma_I(t,\Delta t)  = \left\langle {I}\left(t- \frac{\Delta t}{2}\right) {I}\left(t + \frac{\Delta t}{2}\right) \right\rangle \ .
\label{Ccalt}
\end{equation}

The analogous correlation functions in the frequency domain are given by
\begin{equation}
\widetilde{\Gamma}(\omega, \Delta \omega) = \left\langle \widetilde{E}\left(\omega-\frac{\Delta \omega}{2}\right) \widetilde{E}^*\left(\omega+\frac{\Delta \omega}{2}\right)\right\rangle \ ,
\label{G2def}
\end{equation}
and
\begin{equation}
\widetilde{\Gamma}_I(\omega,\Delta\omega)  = \left\langle \widetilde{I}\left(\omega - \frac{\Delta \omega}{2}\right) \widetilde{I}\left(\omega + \frac{\Delta \omega}{2}\right) \right\rangle \ .
\label{CcalG2}
\end{equation}

It is also customary to define normalized correlation functions~\cite{Saldin1998}. For example in the frequency domain:
%both in time
%
%\begin{align}
%{g}_1(t, \Delta t) 
%&&= \frac{\left\langle {E}(t-\Delta t/2) {E}^*(t+\Delta t/2)\right\rangle}{\left\langle \left|{E}(t-\Delta t/2)\right|^2\right\rangle^{1/2} \left\langle\left|{E}(t+\Delta t/2)\right|^2\right\rangle^{1/2}} \cr
%&&= \frac{\Gamma(t, \Delta t)}{\sqrt{\left\langle {I}(t-\Delta t/2)\right\rangle \left\langle {I}(t+\Delta t/2)\right\rangle}}
%\label{g1gent}
%\end{align}
%%
%
%\begin{align}
%{g}_2 (t, \Delta t) 
%&&= \frac{\left\langle {I}(t-\Delta t/2) {I}(t+\Delta t/2)\right\rangle}{\left\langle {I}(t-\Delta t/2)\right\rangle \left\langle {I}(t+\Delta t/2)\right\rangle } \cr
%&&= \frac{\Gamma_{I}(t, \Delta t)}{\left\langle {I}(t-\Delta t/2)\right\rangle \left\langle {I}(t+\Delta t/2)\right\rangle} \
%\label{g2deft}
%\end{align}

%and the frequency domain

\begin{align}
\widetilde{g}_1(\omega, \Delta \omega) 
&= \frac{\left\langle \widetilde{E}\left(\omega-{\Delta \omega}/{2}\right) \widetilde{E}^*\left(\omega+{\Delta \omega}/{2}\right)\right\rangle}{\left\langle \left|\widetilde{E}\left(\omega-{\Delta \omega}/{2}\right)\right|^2\right\rangle^{1/2} \left\langle\left|\widetilde{E}\left(\omega+{\Delta \omega}/{2}\right)\right|^2\right\rangle^{1/2}} \cr
&= \frac{\widetilde{\Gamma}(\omega, \Delta \omega)}{\sqrt{\left\langle \widetilde{I}\left(\omega-{\Delta \omega}/{2}\right)\right\rangle \left\langle \widetilde{I}\left(\omega+{\Delta \omega}/{2}\right)\right\rangle}}
\label{g1gen}
\end{align}

\begin{align}
\widetilde{g}_2 (\omega, \Delta \omega) 
&= \frac{\left\langle \widetilde{I}\left(\omega-{\Delta \omega}/{2}\right) \widetilde{I}\left(\omega+{\Delta \omega}/{2}\right)\right\rangle}{\left\langle \widetilde{I}\left(\omega-{\Delta \omega}/{2}\right)\right\rangle \left\langle \widetilde{I}\left(\omega+{\Delta \omega}/{2}\right)\right\rangle } \cr
&= \frac{\widetilde{\Gamma}_I(\omega, \Delta \omega)}{\left\langle \widetilde{I}\left(\omega-{\Delta \omega}/{2}\right)\right\rangle \left\langle \widetilde{I}\left(\omega+{\Delta \omega}/{2}\right)\right\rangle}
\label{g2def}
\end{align}

In order to characterize the radiation pulse in both domains, we consider a generalized class of time-frequency signal representations that was introduced by Cohen~\cite{Cohen1966,Cohen1995book}:

\begin{align}
\mathcal{C}(t, \omega) &=
\frac{1}{4\pi^2}\iiint_{-\infty}^{\infty} d(\Delta\omega) d(\Delta t) du\hphantom{x} {\Gamma}(u, \Delta t)\phi(\Delta\omega,\Delta t)\exp(i \Delta\omega t + i \Delta t \omega - i \Delta\omega u) \cr &= \frac{1}{4\pi^2}\iiint_{-\infty}^{\infty} d(\Delta\omega) d(\Delta t) du\hphantom{x} \widetilde{\Gamma}(u, \Delta\omega)\phi(\Delta\omega,\Delta t)\exp(i \Delta\omega t + i \Delta t \omega - i \Delta\omega u)\,
\label{eq:Cohen}
\end{align}
where $ \phi(\Delta\omega,\Delta t) $ is a two-dimensional function called kernel~\cite{Claasen1980} and determines the particular representation in the class.

Wigner distribution is one of the time-frequency representations, useful to describe properties of FEL radiation~\cite{Wu2007,Allaria2010,Marcus2014,Huang2016,Serkez2016_srn}:
\begin{align}
\mathcal{W}(t, \omega) &=
\frac{1}{2\pi}\int_{-\infty}^{\infty} d(\Delta t) {\Gamma}(t, \Delta t)\exp(i \omega  \Delta t) \cr &= \int_{-\infty}^{\infty} d(\Delta\omega) \widetilde{\Gamma}(\omega, \Delta \omega)\exp(- i \Delta\omega t) \ .
\label{eq:Wigner}
\end{align}
% Bastiaans 1997 (inv)

%The Wigner distribution of a signal is a representation in time and frequency domains simultaneously. It amounts to a generalization of the concept of phase-space distribution. % and can be calculated either for a single realization of random process or averaged over a statistical ensemble.
Its kernel $ \phi(\Delta\omega,\Delta t)=1 $. Wigner distribution is real but can take negative values, and yields so called ``cross terms'' between signals in time-frequency plane. Nevertheless, when averaged over an ensemble, is positive for a great number of nonstationary processes which allows one to interpret it similarly to radiation spectrogram~\cite{Flandrin1986}. Its marginal distributions (projections on time and frequency domains) are radiation power and spectral power correspondingly%~\cite{Bastiaans1986, Bastiaans1997}
:
          
\begin{align}
\left\langle I(t)\right\rangle\ &= \int_{-\infty}^{\infty}\mathcal{W}(t, \omega) d\omega,\cr
\left\langle\widetilde{I}(\omega)\right\rangle &= \int_{-\infty}^{\infty}\mathcal{W}(t, \omega) dt,
\label{eq:Wigner_proj}
\end{align}

Its first conditional moments of time and frequency are the instantaneous frequency and group delay:

\begin{align}
\langle\omega\rangle_t  &= \frac{1}{\left\langle I(t)\right\rangle} \int_{-\infty}^{\infty}\omega\mathcal{W}(t, \omega) d\omega,\cr
\langle t\rangle_\omega &= \frac{1}{\left\langle \widetilde{I}(\omega)\right\rangle} \int_{-\infty}^{\infty}t\mathcal{W}(t, \omega) dt.
\label{eq:Wigner_inst_moments}
\end{align}

The conditional spreads of Wigner distribution in frequency and time can also be introduced as instantaneous bandwidth and group delay spread
%kohen TFanalysis pp.120, 
\begin{align}
\sigma^2_{\omega|t} &= \langle\omega^2\rangle_t - \langle\omega\rangle_t^2 = \frac{1}{\left\langle I(t)\right\rangle} \int_{-\infty}^{\infty}\left(\omega-\langle\omega\rangle_t\right)^2\mathcal{W}(t, \omega) d\omega,\cr
\sigma^2_{t|\omega} &= \langle t^2\rangle_\omega - \langle t\rangle_\omega^2 = \frac{1}{\left\langle \widetilde{I}(\omega)\right\rangle} \int_{-\infty}^{\infty}\left(t-\langle t\rangle_\omega\right)^2\mathcal{W}(t, \omega) dt,
\label{eq:Wigner_2nd_moments}
\end{align}

\noindent alas, they may become negative, hence cannot always be properly interpreted~\cite[p.120]{Cohen1995book}.

Any representation of Cohen class can be expressed in terms of the Wigner distribution:

\begin{align}
\mathcal{C}(t, \omega) &=
\frac{1}{2\pi}\iint_{-\infty}^{\infty} d(\Delta\omega) d(\Delta t) \hphantom{x} \Phi(\Delta t,\Delta\omega) \mathcal{W}(t-\Delta t, \omega-\Delta\omega) \cr &\equiv \frac{1}{2\pi}\Phi(\Delta t,\Delta\omega)** \mathcal{W}(t-\Delta t, \omega-\Delta\omega),
\label{eq:cohen_via_wig}
\end{align}

where 

\begin{equation}
\Phi(t,\omega) = \frac{1}{2\pi}\iint_{-\infty}^{\infty} d(\Delta\omega) d(\Delta t) \phi(\Delta t, \Delta\omega) \exp(- i \Delta\omega t + i \Delta t \omega)
\label{eq:bigkernel}
\end{equation}
is just a different representation of Cohen kernel via its two-dimensional Fourier transform. Symbol $ ** $ denotes a two-dimensional convolution operator.

Spectrogram is another and much more known Cohen class function that facilitates time-frequency analysis:

\begin{equation}
S(t, \omega) = \left\langle \left| \frac{1}{\sqrt{2\pi}} \int_{-\infty}^{\infty} d\tau E(\tau)h(\tau-t) \exp(i\omega\tau) \right|^2 \right\rangle.
\end{equation}

It requires introduction of a window function $ h(t) $ in time and frequency domain
\begin{equation}
H(\omega) = \frac{1}{\sqrt{2\pi}} \int_{-\infty}^{\infty} dt h(t) \exp(i\omega\tau).
\end{equation}

Spectrogram kernel function is more complicated:

\begin{equation}
\phi(\Delta\omega,\Delta t) = \int_{-\infty}^{\infty} du \hphantom{x} h(u-\Delta t/2) h^*(u+\Delta t/2) \exp(i \Delta\omega u).
\end{equation}

From Equations~(\ref{eq:cohen_via_wig}) and~(\ref{eq:bigkernel}) it follows that the spectrogram of the signal $ f $ can be obtained by convolving the Wigner distribution $ W_f $ of that signal with the Wigner distribution $ W_h $ of the spectrogram window function $ h $:

\begin{equation}
S_f(t, \omega) = W_f(t,\omega)**W_h(-t,\omega),
\label{eq:Wigner_conv}
\end{equation}

\noindent in other words spectrogram is, in a way, a ``smeared'' version of the Wigner distribution.
In contrast to a Wigner distribution, a spectrogram is non-negative everywhere, but it fails to yield radiation and spectral powers via its marginal distributions, as they are also ``smeared'' by being convolved with window functions:

\begin{align}
\int_{-\infty}^{\infty}S(t, \omega) d(\omega) &= \left\langle I(t)\right\rangle *|h(-t)|^2 ,\cr
\frac{1}{2\pi}\int_{-\infty}^{\infty}S(t, \omega) d(t) &= \left\langle\widetilde{I}(\omega)\right\rangle*|H(\omega)|^2
\label{eq:Spectrogr_proj}
\end{align}

If $ W_f $ is much ``broader'' than $ W_h $ (namely, the radiation pulse is far from its transform limit, or in other words, there are many spikes is the SASE spectrum), then the effect of ``smearing'' is negligible and both Wigner and spectrogram distributions are nearly indistinguishable and up to certain extent can be referred to interchangeably.

%Wigner distribution represents a signal in time and frequency domains simultaneously, can be calculated either for a single realization of random process or averaged over an ensemble. It is a real function and its projections on time and frequency domains are ensemble-averaged radiation power and spectral power correspondingly~\cite{Bastiaans1986, Bastiaans1997}:

\subsection{Stationarity, Ergodicity and SASE radiation}\label{sec:SASE_process}

The electric field of a SASE FEL pulse as a function of time is a stochastic, or random, process.
Generation of the field starts from the shot noise in an electron beam and yields quasimonochromatic radiation~\cite{Saldin2010book}.
%The nature of SASE defines the type of random process that the pulse field belongs to.

Let us consider a random process $a$ where $ ^ka(t) $ is the value of the $ k $-th realization of the variable $ a $ at time $ t $.
The finite time average of the $ k $-th realization of $ a(t) $ over time $ T $ is defined as

\begin{equation}
\left[ ^k a(t) \right]_T \equiv \frac{1}{T} \int_{t-T/2}^{t+T/2} \,^ka(t')dt' \ .
\end{equation}

The full time-average of $k$-th realization is

\begin{equation}
\overline{^ka} = \lim\limits_{T \rightarrow \infty} \left[ ^k a(t) \right]_T.
\end{equation}

The ensemble average of the process $a$ ig given by
%If \rr{we take $N$ realizations of the process $ a $}, the ensemble average (i.e. an average over all possible realizations) of this process depends only on time $ t $: \rr{I would avoid to underline so much dependencies. You just define two averages. That's it.}

\begin{equation}
\left\langle ^ka(t)\right\rangle \equiv \frac{1}{N} \sum_{k=1}^{N} \vphantom{1}^{k}a(t).
\end{equation}

\begin{figure}[h]
	\centering
	\includegraphics[width=0.7\linewidth]{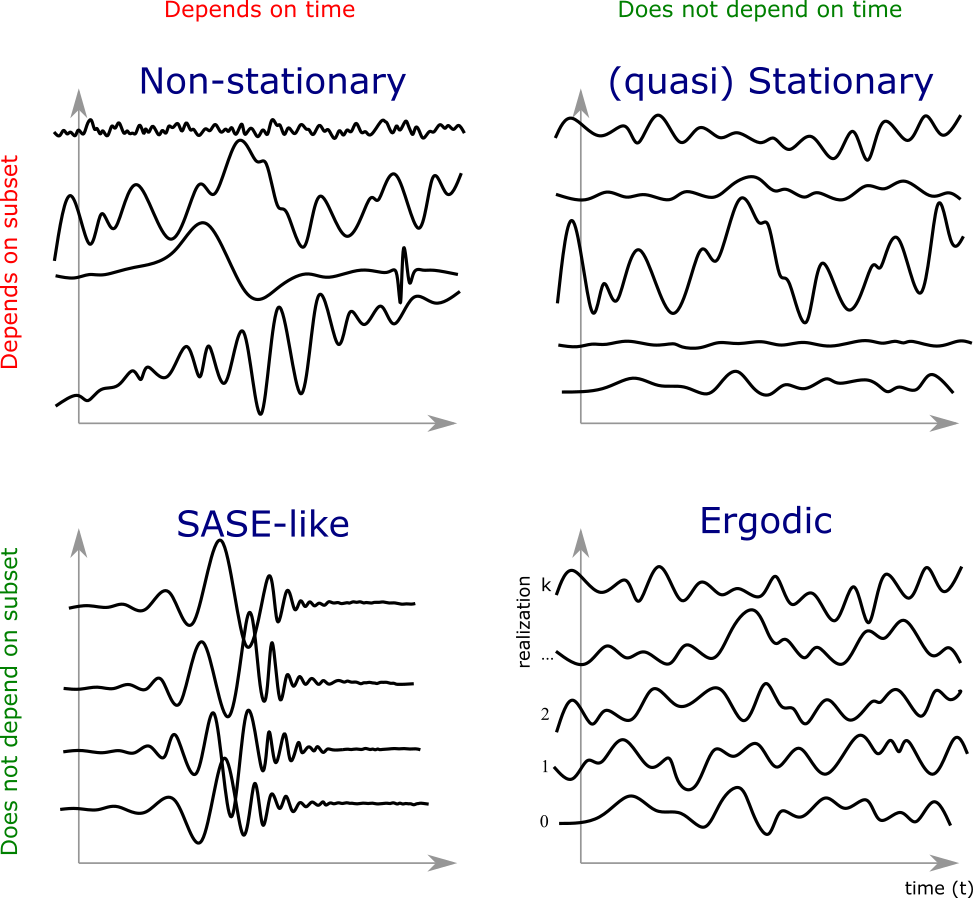}
	\caption{Statistical processes arranged according to whether their average depends on the choice of realization subset or the shift of time.}
	\label{fig:statprocesses}
\end{figure}

We remind that stochastic processes can be separated into several broad classes~\cite{Mandel1995,Goodman2000}, illustrated on Figure~\ref{fig:statprocesses}.
%Statistical properties of the stochastic processes separate them into several broad classes . \rr{Involved.... why not just: "".}

In a \textbf{stationary} random process the probability distribution of any random variable or their combination does not depend on time. 
In a more narrowly defined \textbf{wide-sense stationary} random process, the ensemble-averages $ \left\langle ^ka(t)\right\rangle $ and $ \left\langle ^ka(t) ^ka(t +\Delta t)\right\rangle $ are independent on $t$.

Finally, \textbf{ergodic} random process is the most restrictive subclass of stationary processes. 
It requires that any time average calculated over the sample function must be equal to the same average calculated over the ensemble:

%realization both time and ensemble averages of the random process and all its correlation functions are equal and independent on any subset or time correspondingly: 
\begin{align}
&\overline{^ka} = \langle ^ka(t) \rangle,\cr
&\overline{^ka(t) ^ka(t+\Delta t)} = \langle ^ka(t) ^ka(t+\Delta t)\rangle . \cr
&\textrm{...}
\label{eq:ergodic_def}
\end{align}

It follows from this definition that for ergodic processes the ensemble averages of any combination of random variables are defined and limited.
This is not true for non-stationary or even stationary processes in a general case.

From these definitions it follows that any radiation field described by a stationary random process has no beginning, nor end, which strictly speaking is not physical.
In order to avoid this issue, we call a process \textbf{quasistationary} when the finite time average of each realization $ \left[ ^k a(t) \right]_T $ and its autocorrelation $ \left[ ^ka(t) ^ka(t +\Delta t)\right]_T $ are independent of time $ t $. %\footnote{Provided that the time window $T$ is sufficiently long \rr{..and what means sufficiently long? Maybe better cancel the footnote}}. 
%{\color{red} can't really use coherence time here yet. Have to check whether this is the definition we use later}
%It implies that its normalized autocorrelation (in case of radiation - coherence function $\gamma(t, \Delta t)  = \Gamma(t,\Delta t)/\langle I(t-\Delta t / 2) \rangle \langle I(t + \Delta t /2) \rangle$) is independent of $t$. This can be understood as preservation of coherence time along the pulse. \rr{?}

An important property of \textit{quasistationary} and \textit{ergodic} processes is that their elsemble-averaged Wigner distribution (see Equation~(\ref{eq:Wigner})) can be factorized, and therefore is non-negative~\cite{Flandrin1986}. This property is extremely useful for time-frequency analysis~\cite{Janssen1985}.%.. It allows one to interpret it as time-frequency density distribution~\cite{Flandrin1986, Janssen1985} which is extremely useful for time-frequency analysis.

%\rr{??? See Goodman Figure 3-4, ALSO see (3.2-5)}.

%Position of SASE radiation field in the classification of random processes is illustrated in Figure~\ref{fig:processesdraft}

In principle, SASE FEL radiation in terms of its electric field value as a function of time in the time domain or as a function of frequency in the frequency domain can be categorized within the developed hierarchy of \textit{non-stationary - stationary - ergodic processes} discussed above as a non-stationary process.

SASE radiation however exhibits an important property that distinguishes it from completely generic non-stationary processes: radiation power, instantaneous frequency and instantaneous bandwidth depend on electron beam slice properties, such as beam current, emittance, average and dispersion of electron energy, Twiss parameters, etc. The electron beam, in turn, is usually reproducible on a shot-to-shot basis, which means that the only random process involved in the radiation pulse generation is the intrinsic shot noise in the beam. This specific reproducibility of SASE radiation is illustrated on Figure~\ref{fig:statprocesses}.
%vary slowly along the electron beam with respect to the coherence time, as they depend on electron beam properties, such as beam current, emittance, average and dispersion of electron energy, Twiss parameters, etc. \rr{good?} 
%Note that the electron beam in FELs, in turn, is usually reproducible on a shot-to-shot basis, which means that the only random process involved in the radiation pulse generation is the intrinsic shot noise in the beam. 
%This allows interpretation of the SASE pulses as (finite in time) realizations of the same ensemble, and means that various ensemble averages of radiation quantities, in particular - Wigner distribution, represent the properties of typical radiation pulse upon convergence.\rr{think}

%The electron beam in FELs is usually reproducible on shot-to-shot basis, which leads to reproducible properties of generated radiation (see Figure~\ref{fig:statprocesses}) and allows one to define ensemble-averaged power and spectrum.

Though the SASE generation process is both non-stationary and periodic, through the repetition rate of the electron beam, it should not be confused with \textit{periodic non-stationary} process, discussed in~\cite{Ogura1971, Gardner1975}.
%Here we assume that each SASE pulse is generated by the reasonably similar, reproducible electron beam.

%Therefore, properties of SASE radiation depend on the location along the electron beam, usually vary slowly compared to the coherence time and do not depend on subset of ensemble from which they are calculated.
% \rr{Counter-example: SASE3, 20pC, 5nm}.
%Similarly in the frequency domain, the ensemble-averaged SASE spectrum does not depend on ensemble subset.
% aside from the expected standard error \rr{Average is average, standard error is standard error. Discuss.} inversely proportional to the square root of the sample size.

\begin{figure}[h]
	\centering
	\includegraphics[width=0.45\linewidth]{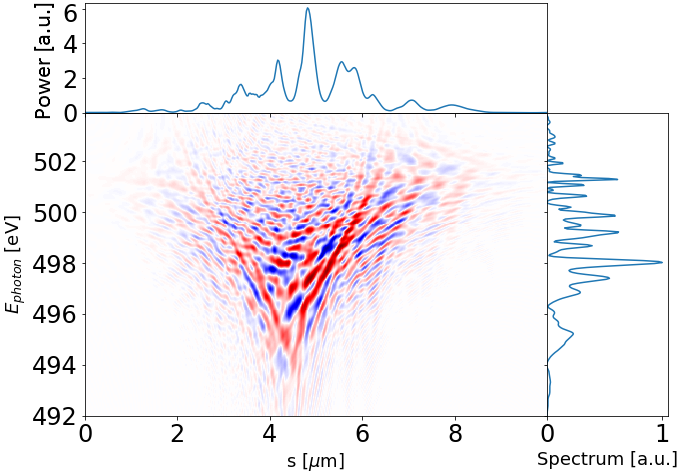}
	\includegraphics[width=0.45\linewidth]{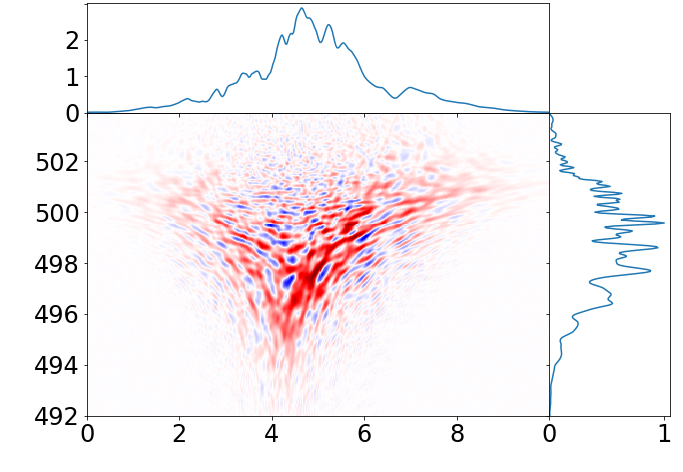}
	\includegraphics[width=0.45\linewidth]{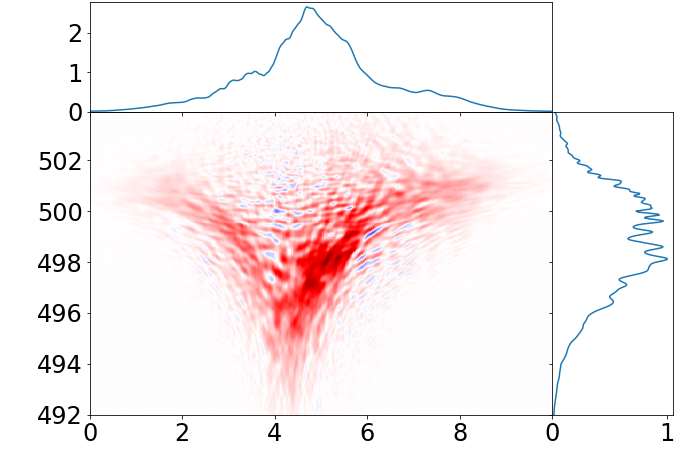}
	\includegraphics[width=0.45\linewidth]{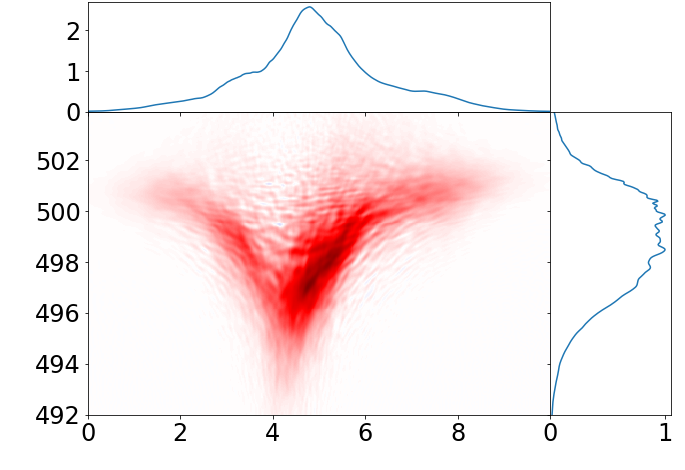}
	\put(5,20){\includegraphics[width=0.05\linewidth]{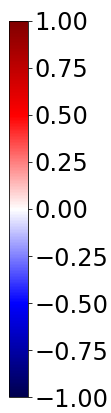}}
	\caption{Colormap representations of the Wigner distribution of simulated SASE FEL radiation with their marginal distributions when averaged over an ensemble of one (upper left subfigure), 10 (upper right), 100 (lower left) and 1000 (lower right) statistically independent realizations. Note the significant non-linear frequency chirp in the pulse, visible upon ensemble averaging. Hereafter the diverging colormap of a Wigner distribution is normalized to its maximum absolute value, while its zero value is depicted with a white color.}
	\label{fig:wig_converge}
\end{figure}

\begin{figure}[h]
	\centering
	\includegraphics[width=0.45\linewidth]{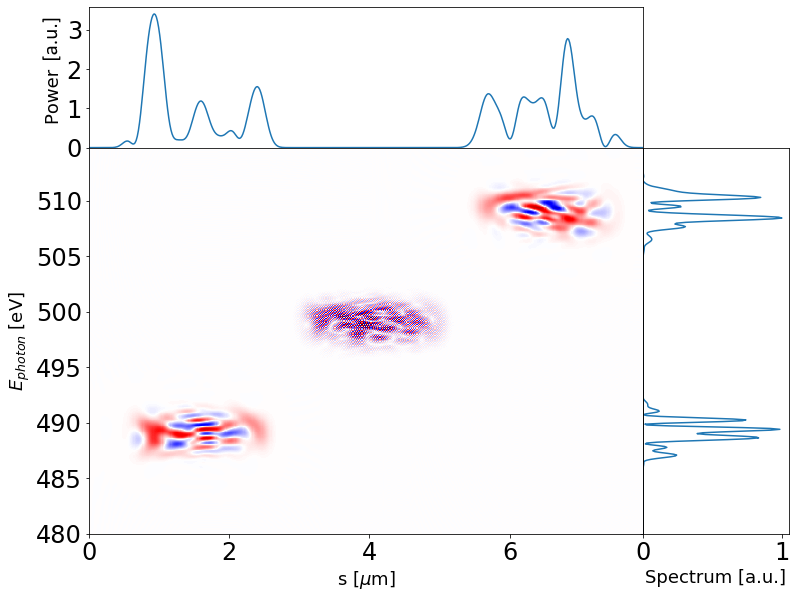}
	\includegraphics[width=0.45\linewidth]{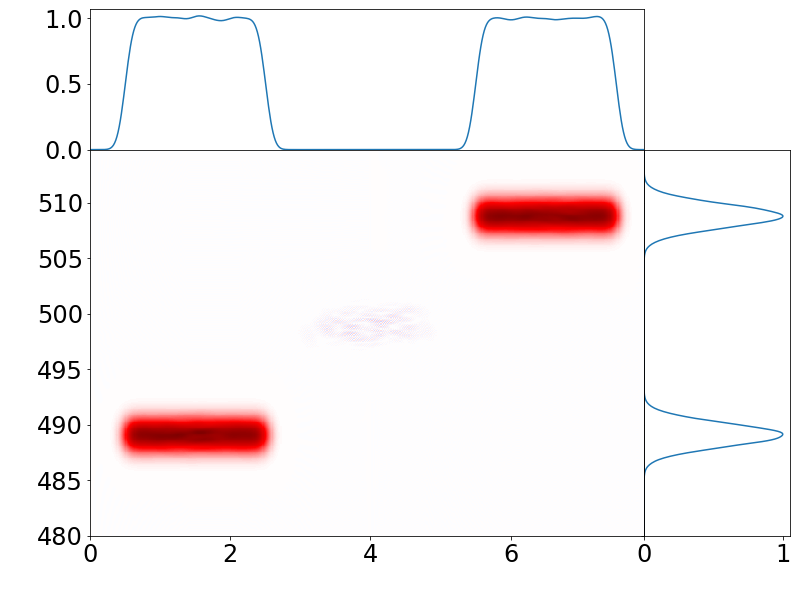}
	\includegraphics[width=0.45\linewidth]{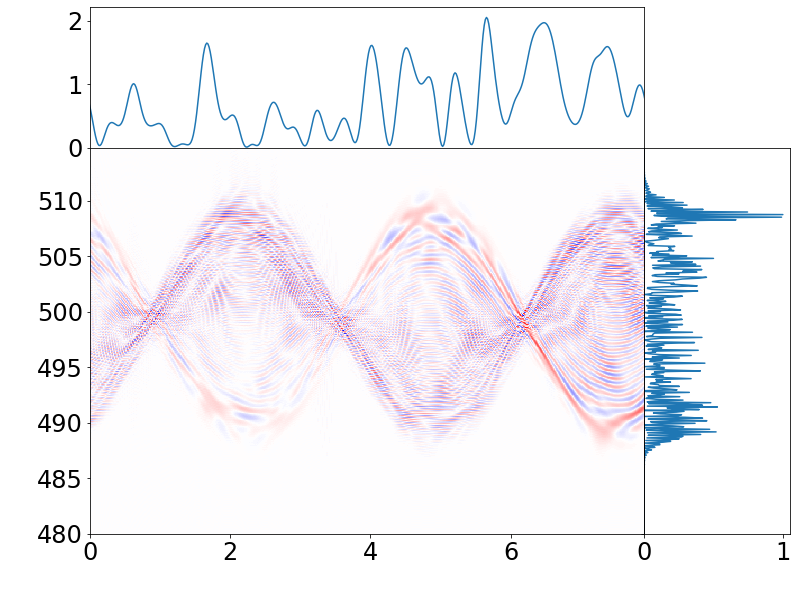}
	\includegraphics[width=0.45\linewidth]{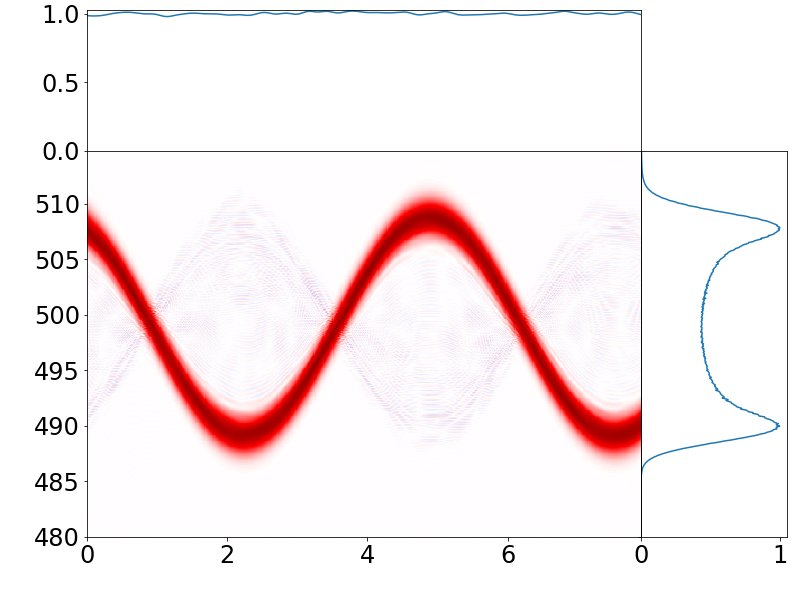}
	
	\caption{Colormap representations of the Wigner distribution of SASE FEL radiation imitation for two "flat-top" pulses with different frequencies (top subfigures) and for a continuous pulse with instantaneous frequency varying in time sinusoidally (bottom subfigures). The distributions averaged over an ensemble of one and 10000 statistically independent realizations are presented on the left and right subfigures correspondingly. The amplitude of the cross-terms is reduced significantly upon averaging over an ensemble.}
	\label{fig:wig_converge_model}
\end{figure}

With examples from~\cite{Flandrin1986} in mind, we find empirically that the ensemble-averaged Wigner distribution of SASE FEL radiation pulses converges towards non-negative values (for an increasing number of realizations in ensemble) even when the Wigner distribution is not separable and not delineated in the time-frequency plane. This convergence is illustrated on Figures~\ref{fig:wig_converge} and~\ref{fig:wig_converge_model} where we present Wigner distributions of SASE FEL pulses, simulated with GENESIS code~\cite{GENE} and of their imitation, modeled with an algorithm presented in~\cite{Pfeifer2010}. Postprocessing is carried out with OCELOT package~\cite{ocelot-collab}. In both cases the radiation has statistical properties of a Gaussian random process~\cite[sec.~3.1.4]{Mandel1995}.

In contrast to convolution with window function to calculate non-negative spectrogram for every event, statistical averaging preserves correct marginals at the cost of losing single-shot information.

%out_stat = read_out_file_stat('/gpfs/exfel/data/scratch/svitozar/projects/XFEL/2017_03_FEL_spectra/20pC_500_fix/new1/', stage=1, run_inp=[])
%wig = wigner_stat(out_stat, pad=2, z=60)
% W = np.mean(wig.wig_stat[0:5],axis=0);print(np.amin(W)/np.amax(W)); print(np.sum(W[W<0]) / np.sum(W[W>0]))
%plt.plot([1,10,100,300,700,1000], [-0.859, -0.601, -0.473,-0.325,-0.152,-0.09])
%plt.plot([1,10,100,300,700,1000], [-0.775,-0.455,-0.137,-0.06,-0.037,-0.0241])
%plt.ylim([-1,0])
%plt.xlim([-10,1000])
%wig.s -= 2.5e-6
%plot_wigner(wig, x_lim=[0,9.98], y_lim=[492,503.9])

%Properties of SASE FEL radiation being both periodical (as its ensemble-averaged Wigner distribution is reproducible) and non-stationary should not be confused with \textit{Periodic Non-stationary} process, discussed in~\cite{Ogura1971, Gardner1975}.

%The shot-to-shot reproducibility of an electron beam parameters allows us to define and calculate ensemble averages introduced in Section~\ref{sec:definitions} for SASE FEL radiation.

%As a result, based on its statistical properties, SASE radiation belongs to a \textbf{Cyclostationary} class of random processes, also known as \textbf{periodic Nonstationary}~\cite{Gardner1975}. See Figure~\ref{fig:processesdraft}. \rr{I think I disagree. Cyclost. means $\langle E(t+\Delta t/2) E^*(t-\Delta t/2) = I(t) \gamma(\Delta t) \rangle$ with I(t) PERIODIC, see Flandrin. You just want that $I(t)$ is a slow function with respect to $\gamma$, I believe! }
%SASE radiation in general may also be chirped, as illustrated on Figure~\ref{fig:processesdraft}

\subsection{Derivation of Wigner Distribution Autocorrelation}

%The amplitude of the radiation field may be very challenging to measure directly. Instead, its spectral density measured and analyzed.

If we assume that scalar field $ E(t) $ obeys Gaussian statistics, 
%\footnote{strictly speaking, only true in a linear regime of operation}, 
then the moment theorem for Gaussian random variables can be applied to the intensity autocorrelation function~\cite[sec.~8.4.1]{Mandel1995} to obtain:
\begin{equation}
\widetilde{\Gamma}_I(\omega,\Delta\omega)  = \left\langle \widetilde{I}\left(\omega - \frac{\Delta \omega}{2}\right)\right\rangle\left\langle \widetilde{I}\left(\omega + \frac{\Delta \omega}{2}\right)\right\rangle + \left|\widetilde{\Gamma}(\omega, \Delta \omega)\right|^2\ .
\label{G2mom}
\end{equation}
%
%In the following, we will refer to $\left|\widetilde{\Gamma}(\omega, \Delta \omega)\right|^2$ as $C(\omega,\Delta\omega)$.
The latter is equivalent to the Siegert relation for normalized correlation functions

\begin{equation}
\widetilde{g}_2 (\omega, \Delta \omega) = 1+ |\widetilde{g}_1 (\omega, \Delta \omega)|^2.
\label{g2Siegert}
\end{equation}

This relation is valid for SASE radiation in both linear amplification regime and saturation~\cite{Saldin1998, Lutman2012}. 
Let us now consider the following Fourier transform:
\begin{equation}
R(t,\omega) = \int_{-\infty}^{\infty} d(\Delta\omega) \left|\widetilde{\Gamma}(\omega, \Delta \omega)\right|^2 \exp(- i \Delta\omega t) \ .
\label{eq:FTR}
\end{equation}

Using the autocorrelation theorem
%\footnote{since the FEL pulses are finite, wide-sense stationarity is not required \rr{?}}, 
we can equate
\begin{align}
&\int_{-\infty}^{\infty} d(\Delta\omega) \left|\widetilde{\Gamma}(\omega, \Delta \omega)\right|^2  \exp(- i \Delta\omega t) \cr & =  \mathcal{A}\left[\int_{-\infty}^{\infty} d(\Delta\omega) \widetilde{\Gamma}(\omega, \Delta \omega)\exp(- i \Delta\omega t)\right]\ ,\cr
\label{eq:Autoc2}
\end{align}
where $ \mathcal{A} $ denotes autocorrelation, defined as it is generally done in signal processing\footnote{Statistical definition of the autocorrelation function is different}:
\begin{align}
\mathcal{A}[f(t)] \equiv \int_{-\infty}^{\infty} f^*(\tau) f(t+\tau) d \tau \ .
\label{autodef}
\end{align}

We now note that the argument of the autocorrelation product in Equation~(\ref{eq:Autoc2}) is the Wigner distribution function (which always has real values), and therefore
\begin{align}
R(t,\omega) = \mathcal{A}\left[\mathcal{W}(t, \omega)\right] = \int_{-\infty}^{\infty} d \tau \mathcal{W}(\tau, \omega) \mathcal{W}(t+\tau, \omega) \ .
\label{final2}
\end{align}

$ R(t,\omega) $ is the autocorrelation of the Wigner distribution and can be directly calculated based on measured spectra of SASE FEL radiation. As discussed above, a Wigner distribution of such radiation becomes mostly positive upon averaging over sufficiently large statistical ensemble. Then it resembles a more known spectrogram distribution and can be analyzed accordingly.
Hereinafter we refer to this function as the Reconstruction of Spectrogram Autocorrelation or for short, ROSA.%\rr{clarify}
%Wigner Function Autocorrelation Reconstruction, or for short, WARP.
%

Another possible representation of the ROSA distribution for Gaussian wavefields can be derived:
%\footnote{Derivation follows derivation of correlation property for Wigner distributions defined without ensemble averages}:

\begin{align}
	R(t,\omega) = \mathcal{W}_y(t, \omega) - \int_{-\infty}^{\infty} d (\Delta t) \exp(i \omega \Delta t) \left\langle A\left[E(t-\Delta t/2)\right]\right\rangle \left\langle A^*\left[E(t+\Delta t/2)\right]\right\rangle  \ ,
	\label{eq:R_expr}
\end{align}
where
\begin{align}
	y(t) = \int_{-\infty}^{\infty} E^*(\tau) E(t+\tau) d \tau = A[E(t)] \ .
	\label{ydef}
\end{align}

ROSA distribution carries information on changes in temporal structure with frequency and vice versa.
%To illustrate this, consider one of the projection properties of $R(t,\omega)$:
%
One of its properties is that its marginal distribution on the frequency domain yields the square of the average radiation spectrum.
\begin{align}
\int_{-\infty}^{\infty} dt R(t,\omega) &= \int_{-\infty}^{\infty} d \tau \mathcal{W}(\tau, \omega) 	\left[\int_{-\infty}^{\infty} dt \mathcal{W}(t+\tau, \omega) \right] \cr & = \int_{-\infty}^{\infty} d \tau \mathcal{W}(\tau, \omega) \langle \widetilde{I}(\omega)\rangle = 4\pi^2 \langle \widetilde{I}(\omega) \rangle^2 \ ;
\label{eq:margR_t}
\end{align}
Projection in another direction can be derived:
\begin{align}
	\int_{-\infty}^{\infty} d\omega R(t,\omega) &= \int_{-\infty}^{\infty} d\omega \mathcal{W}_y(t, \omega) \cr &- \int_{-\infty}^{\infty} d\omega \exp(i \omega \Delta t) \int_{-\infty}^{\infty} d (\Delta t) \left\langle A\left[E(t-\Delta t/2)\right]\right\rangle \left\langle A^*\left[E(t+\Delta t/2)\right]\right\rangle \cr & = 2\pi\left[\left\langle \bigl|A\left[E(t)\right]\bigr|^2\right\rangle-\left\langle \bigl|A\left[E(t)\right]\bigr|\right\rangle^2\right] \equiv 2\pi\sigma^2\left[\bigl|A\left[E(t)\right]\bigr|\right] \ ,
	\label{eq:margR_om}
\end{align}
where $\sigma^2[f(t)]$ is variance of a function f. 
A cut $R(t,\omega=const)$ therefore gives a certain correlation-like function of the field after bandpass filter at frequency $ \omega $.

%
%
%\begin{align}
%&&\int_{-\infty}^{\infty} d\omega R(t,\omega) = \left\langle |y(t)|^2 \right\rangle = \left\langle \left|\int_{-\infty}^{\infty}d \tau E^*(\tau)E(t-\tau)\right|^2\right\rangle \ .
%\textnormal{\rr{highly doubt!}}
%\label{eq:margR_omega}
%\end{align}
%
%It can be concluded from these projection properties that a cut $R(t=const,\omega)$ roughly gives a squared ``instantaneous'' spectrum, and a cut $R(t,\omega=const)$ gives a certain correlation-like function of the field.

\subsection{ROSA Algorithm}

The algorithm of reconstruction of the spectrogram autocorrelation is very simple and consists of the following conceptual steps.

First, sufficiently large statistics of single shot SASE FEL spectra, in the form of Eq.~(\ref{singlespec}) is acquired. Here we assume that only SASE-related fluctuations are present. Otherwise, the measured data should be filtered since they are prone to additional jitter, unrelated to the SASE process.

Second, we calculate the quantity

\begin{align}
Q(\omega,\Delta\omega) &\equiv \left|\widetilde{\Gamma}(\omega, \Delta \omega)\right|^2 \cr  &= \left\langle \widetilde{I}\left(\omega - \frac{\Delta \omega}{2}\right) \widetilde{I}\left(\omega + \frac{\Delta \omega}{2}\right) \right\rangle - \left\langle \widetilde{I}\left(\omega - \frac{\Delta \omega}{2}\right) \right\rangle \left\langle \widetilde{I}\left(\omega + \frac{\Delta \omega}{2}\right) \right\rangle ~.
\label{Ccal}
\end{align}

%where the brackets $\langle ... \rangle$ indicate, again, ensemble average. 

Finally, third, an inverse Fourier transform yields the reconstruction function $ R $:

\begin{align}
R(t,\omega) = \int_{-\infty}^{\infty} d(\Delta\omega) Q(\omega,\Delta\omega) \exp(- i \Delta\omega t)~.
\label{eq:FTR1}
\end{align}

We have found that binning of reconstruction function $ R $ over several points in both dimensions greatly reduces numerical noise with practically no cost  for effective resolution. This binning effectively serves as convolution of Wigner distribution of the signal with that of a window function, as in Equation~(\ref{eq:Wigner_conv}), with its consecutive desampling.

%
%We now fix a certain value $\omega = \omega_0$, and consider the quantity $R(t, \omega_0)$. This function is the autocorrelation of the FEL pulse seen, in the time domain, through a monochromator with central frequency $\omega_0$ and bandwidth $\sigma_\omega$ large enough so that if we call with $\sigma_t$ the rms pulse duration reconstructed from $R(t, \omega_0)$, one has $\sigma_t \sigma_\omega > \mathrm{TBP}$, with TBP the relevant time-bandwidth product, i.e. the time-bandwidth product of the pulse seen through the monochromator. In other words, the monochromator bandwidth should be large enough so that the pulse through the monochromator is nearly, but not quite bandwidth-limited. This clarifies the physical meaning of ``pulse duration at frequency $\omega_0$'', which is mathematically defined by $\sigma_t$. \rr{reformulate?}

%In the following we will show substantiate this claim, showing how the quantity $R$ is closely related to the pulse duration. We will first consider a simple derivation based on the assumption of quasi-stationarity and on the Sievert relation, meaning that we are in the linear regime where the FEL process follows Gaussian statistics. Further on we will relax the assumption on quasi-stationarity.

\subsection{Factorization of quasistationary pulses}

Let us see what simplifications arise if the considered Gaussian process is also quasistationary. 
The assumption of quasi-stationarity allows us to neglect the dependence of the normalized correlation functions on central frequency $\omega$ i.e. they depend only on frequency separation: $ g_1(\Delta\omega), g_2(\Delta\omega)$.

This allows one to factorize autocorrelation functions

\begin{align}
\widetilde{\Gamma}(\omega, \Delta \omega) &= \left\langle \widetilde{I}(\omega) \right\rangle  \widetilde{g}_1 (\Delta \omega)~,\\
\widetilde{\Gamma}_I(\omega, \Delta \omega) &= \left\langle \widetilde{I}(\omega) \right\rangle^2  \widetilde{g}_2 (\Delta \omega)
\label{qsG}
\end{align}

and, consequently the reconstruction function,

\begin{align}
R(t,\omega) =  {\left\langle \widetilde{I}\left(\omega \right) \right\rangle^2} \mathcal{A}\left[\langle I(t) \rangle \right]~.
\label{eq:factored}
\end{align}
By integrating over frequencies, or fixing any value  $\omega = \omega_0$, we see that the reconstructed function $\int_{-\infty}^{\infty} R(t,\omega) d\omega$ or, alternatively, $R(t,\omega_0)$ is simply, aside for an unimportant multiplicative constant, the autocorrelation of the FEL pulse power in the time domain. %For the previously-made illustration when  $\langle I(t) \rangle$ is a square function, we obtain a triangle function.

\subsection{Analogy with the Wiener-Khinchin theorem}

Here we discuss an interesting analogy between Equation~(\ref{eq:FTR}) and the famous Wiener-Khinchin theorem. The Wiener-Khinchin theorem plays a central role in Fourier-transform spectroscopy and allows one to calculate the radiation spectral density after directly measuring the amplitude autocorrelation function of the radiation in the time domain. We remind that the this theorem holds for wide-sense stationary processes and can be formulated, mathematically, as:
%\rr{arg1}One can see that the Equation~(\ref{eq:FTR}) is very similar to the formulation of the Wiener-Khinchin theorem:
\begin{equation}
\langle\widetilde{I}(\omega)\rangle = \frac{1}{2\pi} \int_{-\infty}^{\infty} d(\Delta t)~ \Gamma(\Delta t) \exp(i \omega \Delta t)~.
\label{eq:W-Kh}
%Mandel-Wolf Eq. 2.4-37, 2.4-15
\end{equation}

%The analogy holds except for the squared modulus of the autocorrelation function and for the opposite direction of the Fourier transform. \rr{These are big differences though. How about showing what happens by applying a slightly improved version of WK -and its opposite- for QS sources? It was there in the previous versions, but not in this one. This would explain the analogy. Otherwise, we are only stating without explanations and better to just leave it out...}

It states that the statistical autocorrelation function of a stationary random process in the \textit{time} domain and the ensemble-averaged spectral density, i.e. power spectrum, of that process form a Fourier transform pair. 

Should the process be quasi-stationary, a similar relation would naturally hold between the autocorrelation function in the \textit{frequency} domain and the ensemble-averaged intensity, or pulse shape, in the time domain: 
\begin{equation}
\langle I(t)\rangle = \int_{-\infty}^{\infty} d(\Delta\omega)~ \widetilde{\Gamma}(\Delta \omega) \exp(-i \Delta\omega t)~.
\label{eq:iW-Kh}
%Mandel-Wolf Eq. 2.4-37, 2.4-15
\end{equation}

One can now see the similarity between Equations~(\ref{eq:iW-Kh}) and (\ref{eq:FTR}). 
%%Let us now consider Equation~(\ref{eq:FTR}). There we calculate the \textit{square modulus} of the statistical autocorrelation function of a \textit{non-stationary} process in the \textit{frequency} domain and carry out the Fourier transform to the time domain in order to obtain the reconstruction function $ R(t,\omega) $. it looks completely different from Equation~(\ref{eq:W-Kh}). However, should the considered process be stationary, the function $ R $ would be factorizable, and the autocorrelation of its \textit{ensemble-averaged} power profile could be reconstructed (Equation~(\ref{eq:factored})). 
The latter is simply the combination of the autocorrelation theorem and the Wiener-Khinchin theorem applied in the opposite domain.

%The Wiener-Khinchin theorem requires that the statistical process is stationary, i.e. \textit{unlimited} in the \textit{time domain} and applies Fourier transform from the time to the frequency domain. %This is not the case for the SASE radiation in its frequency domain, bandwidth-limited frequency domain of
%In our case, we measure the square modulus of the radiation field amplitude in the \textit{frequency domain}, signal in which is \textit{bandwidth-limited}, calculate its autocorrelation function and, similarly to Equation~(\ref{eq:W-Kh}), apply a Fourier transform, but from the frequency to the time domain.

%In Fourier-transform spectroscopy the investigated process is assumed to be ergodic where statistical autocorrelation function is equal to the time autocorrelation function, see Equation~(\ref{eq:ergodic_def}).

In Fourier-transform spectroscopy the \textit{time} autocorrelation function $ \Gamma $ is obtained by carrying average over time, as the investigated process is assumed to be ergodic. This averaging naturally leads to the loss of the dependence of the function $ \Gamma $ from time, so one obtains the statistical autocorrelation function of time delay $ \Gamma(\Delta t) $ over which the Fourier transform is later performed.

In our case, the \textit{statistical} frequency autocorrelation function $ \widetilde{\Gamma}(\omega, \Delta \omega) $ is calculated via averaging over an ensemble of single-shot spectra. Such averaging is justified by statistical properties of SASE radiation, discussed in Section~\ref{sec:SASE_process}.
Dependence on both central frequency $ \omega $ and frequency difference $ \Delta\omega $ remains, yielding additional information upon Fourier transform over $ \Delta\omega $.

%The Wiener-Khinchin theorem usage in spectroscopy,  the autocorrelation function  is obtained not by averaging over time  the product of spectral density only over the statistical ensemble, we have the possibility to study the autocorrelation as a function of central frequency. 

\section{Numerical simulations and discussions}\label{sec:simulations}

In order to illustrate the performance of ROSA, we simulated four ensembles of FEL spectra with the FEL code GENESIS~\cite{GENE} and analyzed them with the OCELOT package~\cite{ocelot, ocelot-collab}. We generated
500 statistically independent SASE events assuming a model 6$\mu$m-long flattop electron beams with (i) and without (ii) energy chirp. Also we simulated SASE generation by two 2$\mu$m-long flattop electron beams separated by 3~micrometers (iii). Finally, 1000 statistically independent SASE events assuming a nominal 100pC electron beam from s2e simulations for the European XFEL (iv)~\cite{Altarelli2006,mpy_web}.
%(i) 1000 statistically independent SASE events assuming a nominal 100pC electron beam from s2e simulations for the European XFEL\rr{Cite s2e sims in MPY} as well as 500 events assuming a model 6$\mu$m-long flattop electron beams with (ii) and without (iii) energy chirp. Finally, (iv) we considered another model beam, composed by two 2$\mu$m-long flattop electron beams separated by 3~micrometers. 
The slice properties of the model beams are chosen to be close to those of the 100~pC nominal electron beam. The radiation generated by model beams was dumped before saturation to reduce the radiation slippage and maintain the illustrative flat-top distribution of the ensemble-averaged radiation power. The realistic 100~pC beam radiation was dumped, instead, in deep saturation.

\begin{figure}
	\centering
	\includegraphics[height=3.7cm]{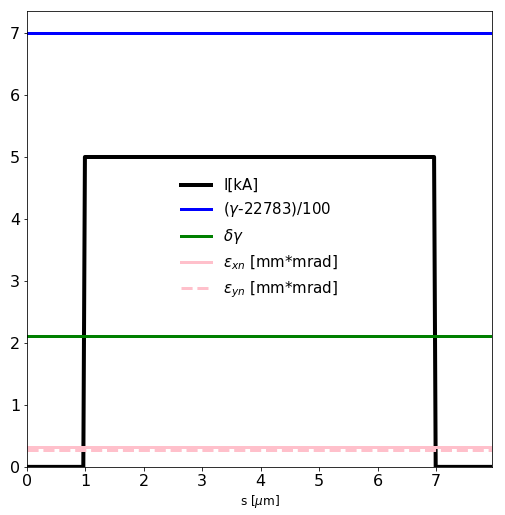}
	\includegraphics[height=3.7cm]{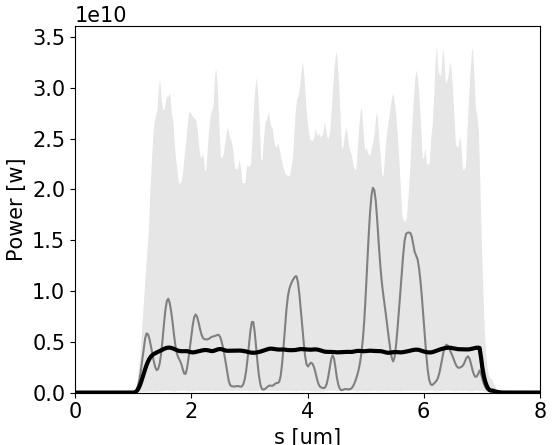}
	\includegraphics[height=3.7cm]{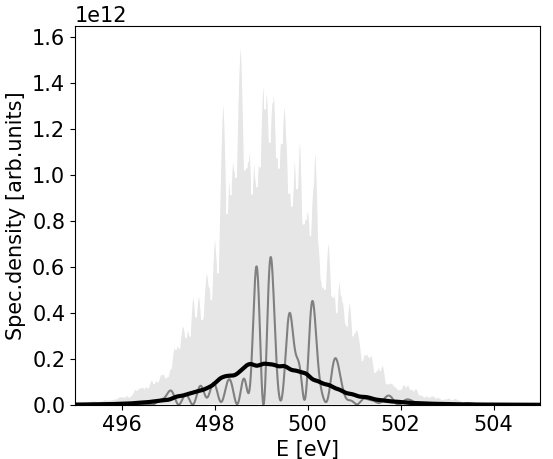}
	\includegraphics[height=6cm]{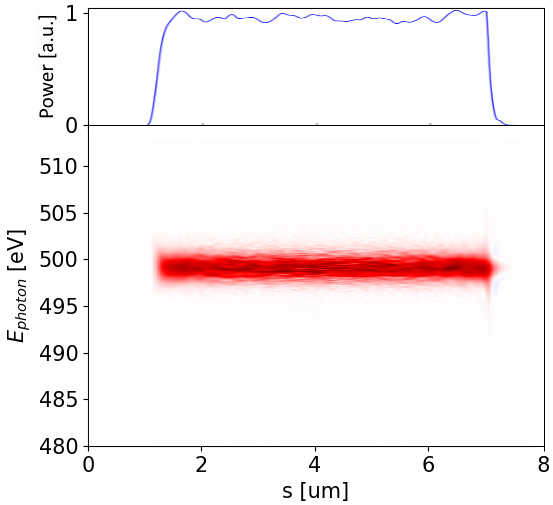}
	\includegraphics[height=6cm]{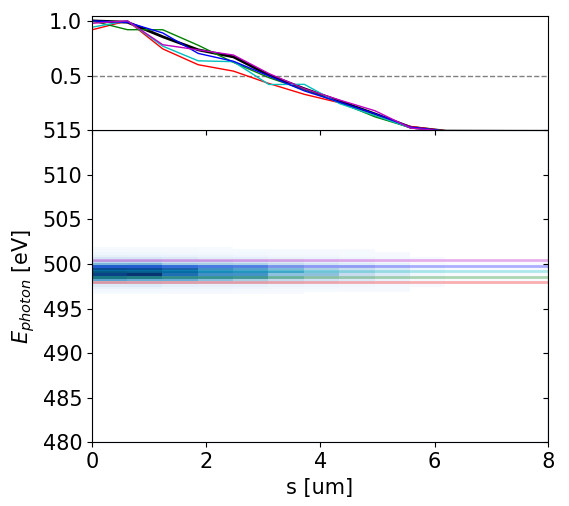}
	\caption{A 6$ \mu $m-long flattop model electron beam without energy chirp, see top left plot, was used to generate SASE radiation. It was dumped during the exponential growth. Power and spectra of 500 statistically independent events are presented on the middle and on the right top plots correspondingly. The ensemble-averaged Wigner distribution of the SASE radiation is presented on the bottom left plot. The spectrogram autocorrelation reconstruction $ R(ct,\hbar\omega/e)$ is presented on the bottom right plot.}	
	\label{fig:6um}
\end{figure}

The bottom left plots in Figures~\ref{fig:6um} and following present the ensemble-averaged Wigner distribution of the radiation. This distribution is based on information about amplitudes and phases of SASE radiation provided by simulation and is used here to assess representativeness of the bottom right plot.
The latter depicts the ROSA function, calculated with Equation~(\ref{eq:FTR}). The colored line-offs allow one to compare the reconstructed spectrogram autocorrelation at the various photon energies. The reconstruction is symmetrical with respect to coordinate $ s $, hence only a half of it is depicted.

If no energy chirp is present in the electron beam, the undulator resonance condition is constant along the beam and the generated radiation pulse has no frequency chirp (Figure~\ref{fig:6um}). In this special case the Wigner distribution, and hence the reconstruction, are factorisable (Equation~(\ref{eq:factored})) and the total pulse length can be estimated: the autocorrelation of the flat-top power profile with length $ \Delta s $ would yield an autocorrelation result with triangular shape and full width at half maximum (fwhm) equal to $ \Delta s / 2 $. In the case of a Gaussian radiation pulse with fwhm $\Delta s$, the fwhm size of its autocorrelation result will be $ \sqrt{2}\Delta s$.

\begin{figure}
	\centering
	\includegraphics[height=3.7cm]{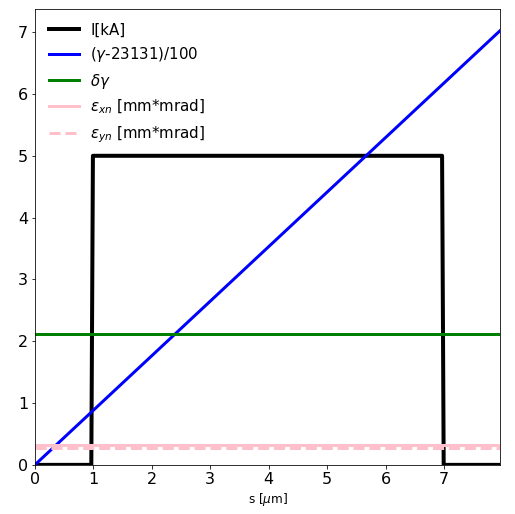}
	\includegraphics[height=3.7cm]{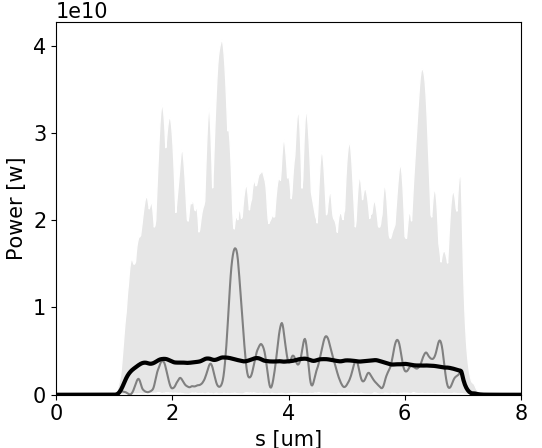}
	\includegraphics[height=3.7cm]{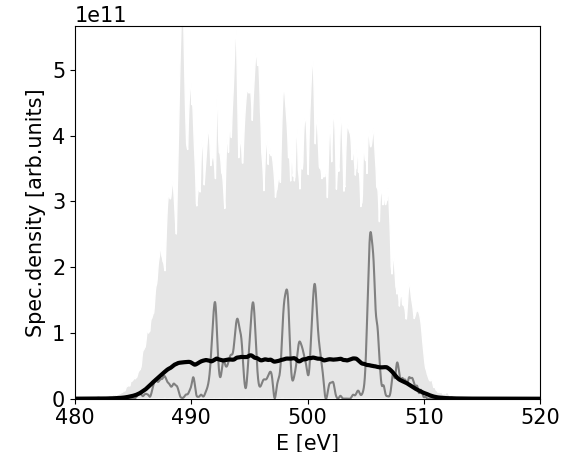}
	\includegraphics[height=6cm]{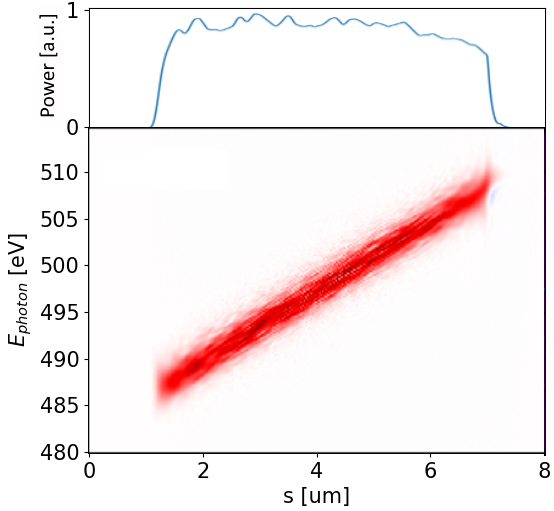}
	\includegraphics[height=6cm]{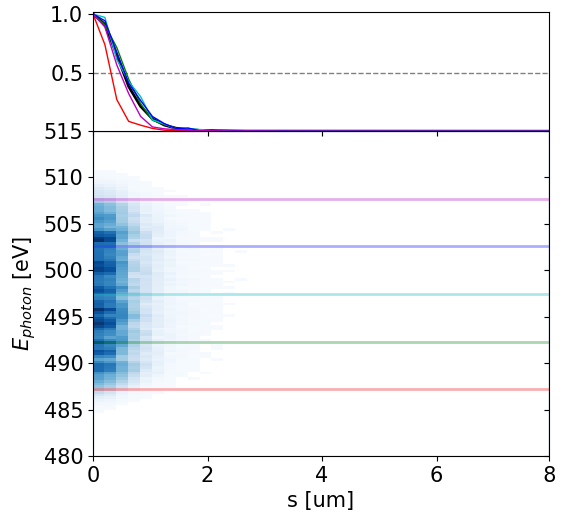}
	\caption{See plots descriptions from Figure~\ref{fig:6um}, except in this case the linear energy chirp is introduced to the electron beam.}
	\label{fig:6um_chirp}
\end{figure}

When the energy chirp in the electron beam (i.e. the relative difference of electron energy in the head and tail) exceeds FEL efficiency parameter  $\Delta\gamma/\gamma\gtrsim\rho $, a frequency chirp along SASE pulse can be observed. As a consequence it will yield a broader spectrum (which is the integral of the Wigner distribution over time) and typically a shorter pulse length at all photon energies~\cite{Krinsky2003} (horizontal line-offs of the Wigner distribution), as presented on Figure~\ref{fig:6um_chirp}. These effects are also reflected in the spectral correlation functions, and, if not accounted for, an underestimation of the total pulse duration will take place.

Two sequential radiation pulses with equal carrier frequencies, separated by time $ \Delta t $ will yield a spectrum with modulated amplitude. The spectral modulation period is given by 
\begin{equation}
 \delta E~[eV] = 10^{6}\frac{hc}{e\Delta s~[\mu m]} \simeq \frac{1.240}{\Delta s~[\mu m]} \simeq \frac{4.135}{\Delta t~[fs]}
 %\newline\\
 %\delta E~[eV] = 10^{15}\frac{h}{e\Delta t~[fs]} \simeq \frac{4.135}{\Delta t~[fs]},
 \nonumber
\end{equation}

%or 
%
%$ \lambda~[eV] = 10^{15}\frac{h}{e\Delta t~[fs]} \simeq \frac{4.135}{\Delta t~[fs]}$,
where $ h $ is the Planck's constant, $ c $ is speed of light, $ e $ is the electron charge, $ s $ and $ t $ are the pulse separations in space (micrometers) and time (femtoseconds) correspondingly. The discussed modulation of spectral density takes place only at the frequencies common for both pulses, therefore if individual spectra of the two pulses do not have common frequencies, i.e. do not overlap in the frequency domain, no ``beating'' in this domain will take place.

\begin{figure}
	\centering
	\includegraphics[height=3.7cm]{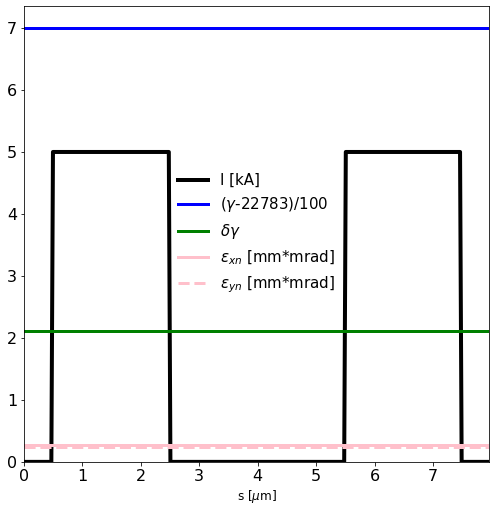}
	\includegraphics[height=3.7cm]{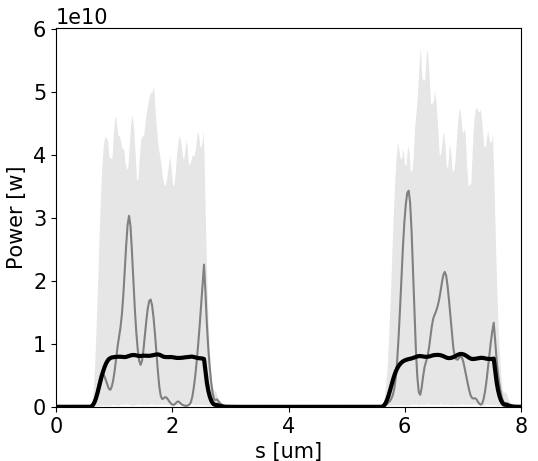}
	\includegraphics[height=3.7cm]{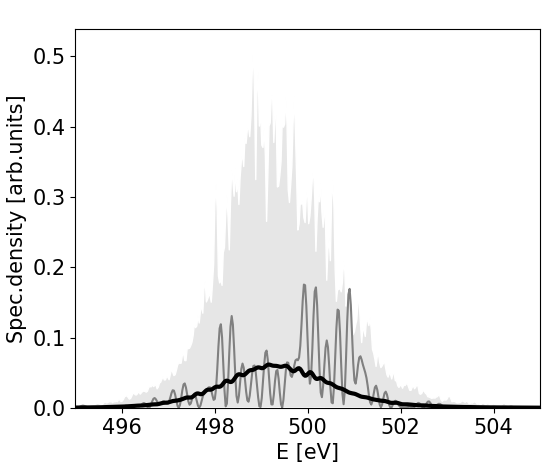}	
	\includegraphics[height=6cm]{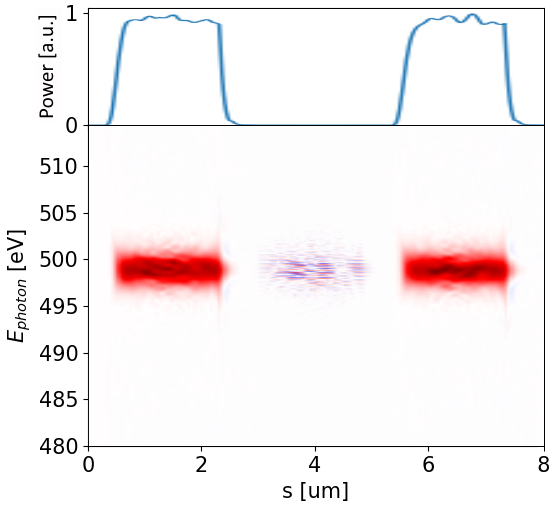}
	\includegraphics[height=6cm]{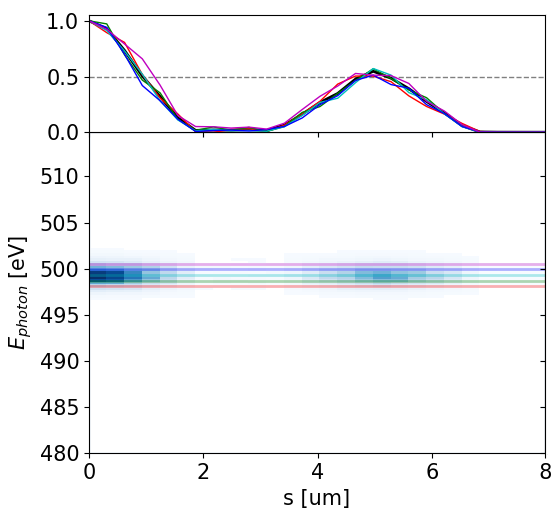}
	\caption{See plots descriptions from Figure~\ref{fig:6um}. Here two flat-top 2~$ \mu $m-long electron beams, separated by 3~$ \mu $m, generate two consecutive SASE pulses of the same averaged shape.}
	\label{fig:2_3_2um}
\end{figure}

\begin{figure}
	\centering
	\includegraphics[height=5cm]{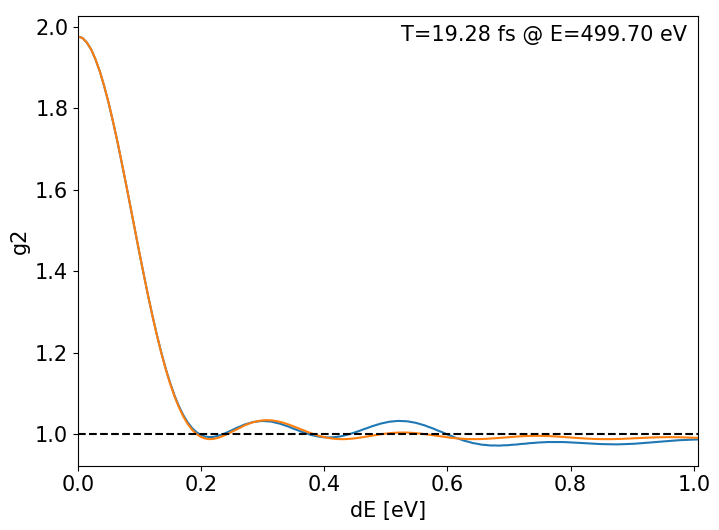}
	\includegraphics[height=5cm]{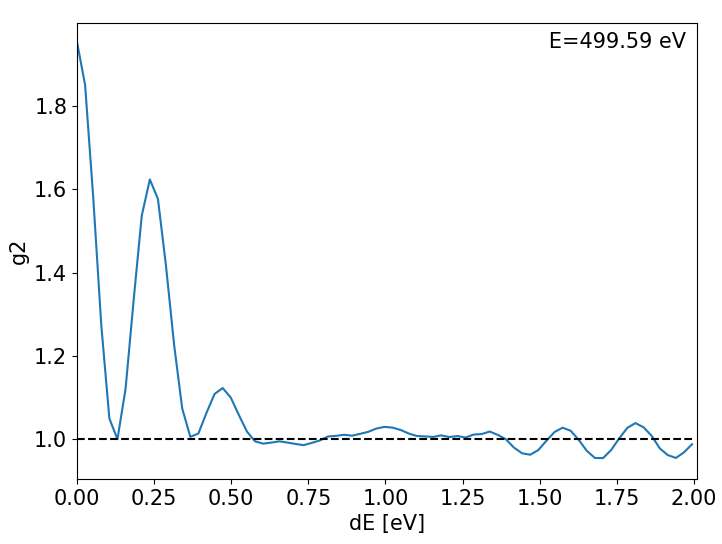}
	\caption{Normalized second-order spectral correlation functions $ g_2 $. The left plot corresponds to the single radiation pulse discussed on Figure~\ref{fig:6um}; the right plot corresponds to the two pulses on Figure~\ref{fig:2_3_2um}. In the case of the 6~$ \mu $m-long radiation pulse with flattop averaged power profile, we fit the function $ g_2 $  with a theoretical shape corresponding to the pulse duration of 19.3~fs.}
	\label{fig:g2_comp}
\end{figure}

Similarly, the SASE spectra obtained from two model flat-top electron beams with equal energies are expected to be modulated. If a spectrometer is capable of resolving this modulation, one can estimate their temporal separation. This scenario is exemplified on Figure~\ref{fig:2_3_2um}. Since the SASE spectra are already intrinsically ``spiky'', it is not easy to discover additional modulation just upon visual examination. However, the spectral correlation function clearly indicates the correlation between the maxima in the spectrum (see Figure~\ref{fig:g2_comp}). The latter does not take place for statistically independent SASE modes.

Note, that both the Wigner distribution and the reconstruction indicate equal temporal separation between SASE pulse centers --- 5~$ \mu $m or 16.5~fs --- at all photon energies.

\begin{figure}
	\centering
	\includegraphics[height=3.7cm]{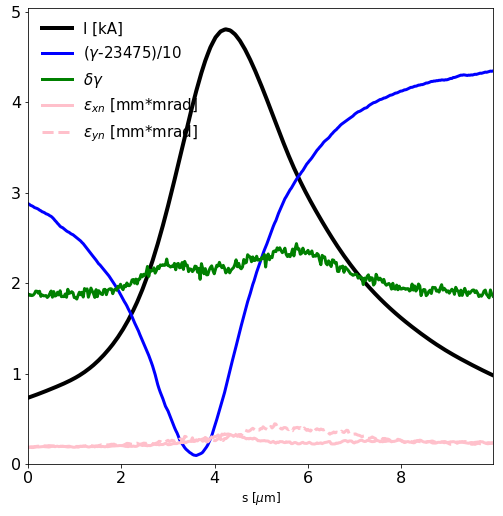}
	\includegraphics[height=3.7cm]{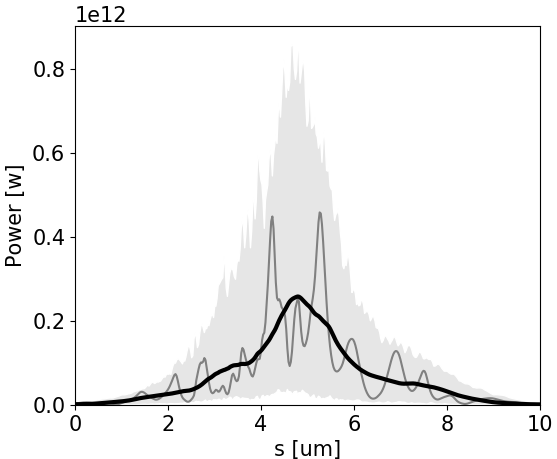}
	\includegraphics[height=3.7cm]{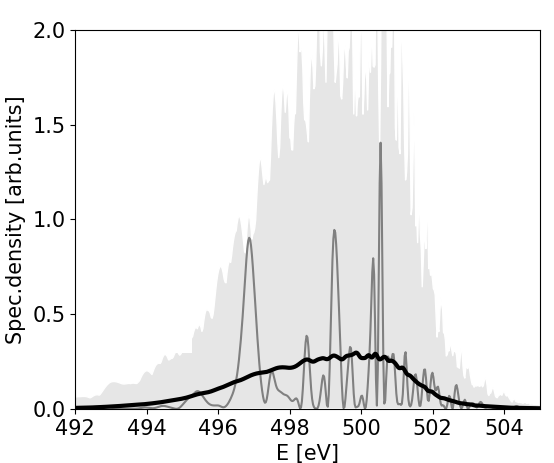}
	\includegraphics[height=6cm]{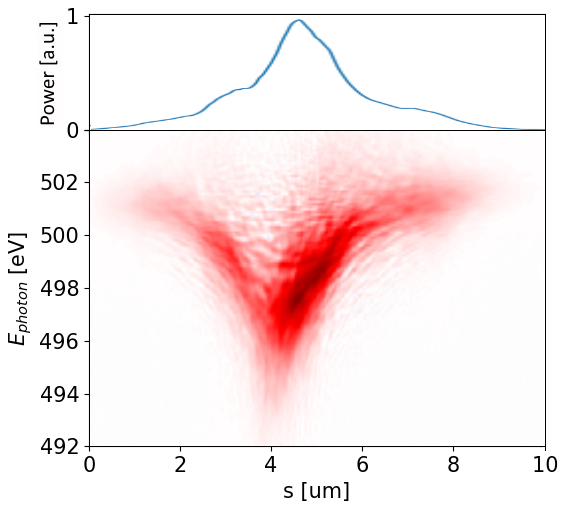}
	\includegraphics[height=6cm]{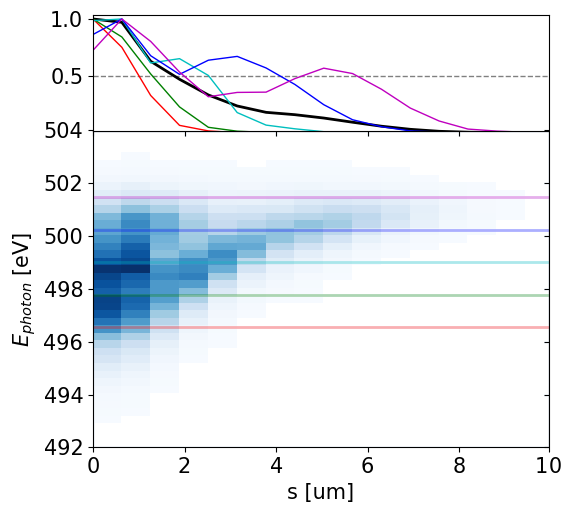}
	\caption{See plots descriptions from Figure~\ref{fig:6um}. The nominal European XFEL 100pC electron beam with a non-linear energy chirp produces SASE radiation with different durations at different photon energies. Note the bifurcation in Wigner distribution above 499~eV. Analysis based on 1000 simulated SASE spectra}
	\label{fig:100pC}
\end{figure}

In general, the electron beam formation system may yield a  highly non-linear energy chirp, as illustrated on Figure~\ref{fig:100pC} (top left plot). If the relative peak-to-valley energy difference in the electron beam is comparable or larger than a Pierce parameter $ \rho $, the electron beam energy chirp will be imprinted into the SASE radiation spectrogram as a radiation frequency chirp. In the given example two distinct pulses separated by about 6~$\mu$m at 501~eV photon energy are visible on the radiation spectrogram. The separation of these ``double pulses'' grows with the photon energy, following the separation of the electron beam slices with an equal Lorenz factor $ \gamma $. Similarly to the double-pulse case, illustrated in Figure~\ref{fig:2_3_2um}, such photon-energy-dependent separation can be straightforwardly observed in the reconstruction function % which is calculated based on the measured SASE FEL specra, 
%and depends on the photon energy \rr{as it should be}.

%\rr{Gaussian statistics in saturation? show re/im in time and freq domain..}

Our spectrum-based pulse length estimation is a zero-cost, effective diagnostics that does not require any hardware aside for a high-resolution single-shot spectrometer, which is typically available at XFEL facilities. It relies upon the fact that the FEL pulses are short, narrow-bandwidth, and follow Gaussian statistics, strictly in the linear regime, and  approximatively at saturation~\cite{Lutman2012}.
The conventional method of fitting the experimental second-order spectral correlation function of SASE radiation with the theoretical one requires an initial assumption on the power profile of the SASE pulse~\cite{Lutman2012, Inubushi2012}. Also, if an energy chirp is present in the electron beam, the fitting will yield an underestimated pulse duration.

We suggest that in many cases, the analysis of a radiation spectrogram by means of its temporal autocorrelation may be more straightforward and less misleading, especially when no information about the electron beam longitudinal phase space is available due to lack of an XTCAV system installed after the SASE undulator.

\section{Conclusions}\label{sec:conclusions}

We discussed several statistical properties of SASE FEL radiation that allow us to define an ensemble-averaged Wigner distribution; we find empirically that it tends to become everywhere positive therefore very useful for time-frequency analysis of SASE FEL radiation.

We present an extend method to analyze the temporal SASE radiation properties. We show that based on the second order spectral correlation function it is possible to calculate a $ R(t, \omega) $ - an autocorrelation of the ensemble-averaged Wigner distribution of the radiation. The latter, upon noise filtering via binning, is close in terms of its properties to the well-known spectrogram distribution. Therefore, we call the proposed method ROSA: Reconstruction of Spectrogram Autocorrelation.

The proposed method allows one to estimate the average group spread in the pulse, i.e. the pulse length individually for all photon energies in the pulse, to indicate the presence of two FEL pulses with overlapping spectra and to estimate their duration and temporal separation. 
This method is statistical in nature and relies upon the assumption that FEL hardware provides a reproducible electron beam along the stable orbit. Otherwise, discrimination of outlier events should take place.

We simulated 500 statistically independent SASE events assuming a model 6$\mu$m-long flattop electron beams with (i) and without (ii) energy chirp. Also we simulated SASE generation by two 2$\mu$m-long flattop electron beams separated by 3~micrometers (iii). Finally, 1000 statistically independent SASE events assuming a nominal 100pC electron beam from s2e simulations for the European XFEL (iv)~\cite{mpy_web}. Simulation results allowed us to compare the apriori known Wigner distributions with calculated ROSA distributions. 

We provided an intuitive understanding of advantages and limitations of ROSA algorithm. Some limitations, like the inability to provide the total pulse duration in the presence of a significant energy chirp in the electron beam without additional knowledge, are common for the all methods based on spectral correlation function analysis.

\section{Acknowledgments}\label{sec:ack}

We would like to thank Guenter Brenner, Stefan Duesterer, Bart Faatz, Jan Gruenert, Vitaly Kocharyan, Naresh Kujala, Jia Liu, Juliane Roensch-Schulenburg, Evgeny Saldin, Takanori Tanikawa, Sergey Tomin, Mikhail Yurkov for useful discussions and Serguei Molodtsov for his interest in this work.

\printbibliography
\end{document}